\documentclass{amsart}
\def\t{\hbox}

\def\f{\frac}

\def\q{\quad}
\def\p{\varphi}

\def\k{\kappa}

\def\a{\alpha}

\def\d{\delta}
\def\be{\begin{equation}}
\def\ee{\end{equation}}
\def\bea{\begin{eqnarray}}
\def\eea{\end{eqnarray}}
\def\ba{\begin{array}}
\def\ea{\end{array}}

\def\pr{\prime}

\newtheorem{theorem}{Theorem}[section]

\theoremstyle{definition}

\theoremstyle{remark}

\numberwithin{equation}{section}

\newcommand{\abs}[1]{\lvert#1\rvert}

\newfont{\Bb}{msbm8 scaled\magstep{1}}
\newcommand{\rc}{\mbox{\Bb R}}

\usepackage{graphics}

\begin{document}

\title
[Fixed-Energy Phase Shifts]
{Piecewise-Constant Positive Potentials with practically the same 
Fixed-Energy Phase Shifts
}


\maketitle
\centerline{\footnotesize
Alexander G. RAMM
\footnote{E-mail: ramm@math.ksu.edu}
}
\vspace{0.3cm}
\centerline{{\footnotesize\it Department of Mathematics, Kansas State University,}}
\centerline{{\footnotesize\it Manhattan, Kansas 66506-2602, U.S.A.}}
\vspace{1cm}
\centerline{\footnotesize
Semion GUTMAN
\footnote{E-mail: sgutman@ou.edu}
}
\vspace{0.3cm}
\centerline{{\footnotesize\it Department of Mathematics, University of Oklahoma,}}
\centerline{{\footnotesize\it Norman, Oklahoma 73019, U.S.A.}}

\vspace{1cm}

{\footnotesize

It has recently been shown that spherically symmetric 
potentials of finite range are uniquely
determined by the part of their phase shifts 
at a fixed energy level
$k^2>0$. However, numerical experiments show that 
 two quite different potentials can produce almost identical phase
shifts. It has been guessed by physicists that such examples are 
possible only for "less physical" oscillating and changing sign
potentials. In this note it is shown that  the above guess
is incorrect:
 we give examples of four positive
spherically symmetric compactly supported quite different
 potentials having practically identical phase shifts.
The note also describes a hybrid 
stochastic-deterministic method for global minimization used for 
the construction of these potentials.

PACS: 0380, 0365.

}

\section{Introduction}
Let $q(x),\, x\in \rc^3$ be a real-valued potential with compact
support. Let $R>0$ be a number
 such that $q(x)=0$ for $\abs{x} > R$. We also
assume that $q\in L^2(B_R)\,,\ B_R=\{x:\abs{x}\leq R, x\in \rc^3\}$. 
Let $S^2$ be the unit sphere, and $\alpha \in S^2$. For
a given energy $k>0$ the scattering solution $\psi(x,\alpha)$ is defined
as the solution of

\begin{equation}
\Delta \psi+k^2\psi-q(x)\psi= 0\,,\quad x \in \rc^3 
\end{equation}
satisfying the radiation condition

\begin{equation}
\lim_{r\rightarrow\infty}\int_{\abs{x}=r}\left| \frac{\partial v}{\partial
r}-ikv\right|^2 ds=0\,,
\end{equation}
where 

\begin{equation}
\psi=\psi_0+v,\quad \psi_0=e^{ik\alpha\cdot x}\,,\quad \alpha\in S^2\,.
\end{equation}

It can be shown, that 

\begin{equation}
\psi(x,\alpha)=\psi_0+A(\alpha',\alpha,k)
\frac{e^{ikr}}{r}+o\left(\frac{1}{r}\right)\,,\;
\text{as}\ \ r\rightarrow\infty\,, \quad
\frac{x}{r}=\alpha '\,.
\end{equation}

The function $A(\alpha',\alpha,k)$ is called 
the scattering amplitude (\cite{n}, \cite{r3}).

For spherically symmetric scatterers $q(x)=q(r)$ the scattering
amplitude satisfies $A(\alpha',\alpha,k)=A(\alpha'\cdot\alpha,k)$. 
The converse is established in
\cite{r6}. Following \cite{rs}, the scattering amplitude
for $q=q(r)$ can be written as

\begin{equation}
A(\alpha',\alpha,k)=\sum^\infty_{l=0}\sum^l_{m=-
l}A_l(k)Y_{lm}(\alpha')\overline{Y_{lm}(\alpha)}\,,
\end{equation}

where $Y_{lm}$ are the spherical harmonics, and the bar denotes the
complex conjugate.

The fixed-energy phase shifts $-\pi<\delta_l\leq\pi$ 
(for this $k$) are defined 
(\cite{rs}) by the formula:

\begin{equation}
A_l(k)=\frac{4\pi}{k}e^{i\delta_l}\sin(\delta_l)\,.
\end{equation}

Thus, every spherically symmetric potential produces a unique set of
phase shifts.
The inverse problem of finding the spherically symmetric potential $q$ by the
set of its phase shifts has long been of interest 
in physics.  Details and applications 
can be found in 
Newton \cite{n} and 
Chadan-Sabatier \cite{cs}, where the authors propose a procedure for the 
identification of the potential $q$. 
This procedure has been recently examined by A. G. Ramm. (see 
[1], part 1).

The following result shows
that, theoretically, the inverse problem 
does have a unique solution given a suitable subset
of single-energy
phase shifts:

\begin{theorem}[ \cite{r5}]
Let ${\mathcal L}$ be an arbitrary fixed subset of non-negative integers
satisfying

\[
\sum_{l\in{\mathcal L},l\neq 0}\frac1l=\infty\,.
\]

Then the data $\{\delta_l\}_{l\in{\mathcal L}}$, corresponding to a
potential $q(r)\in L^2(B_R)\,, R>0\,, r=\abs{x}$, 
$q=0$ for $r>R$, determine $q(r)$
uniquely.
\end{theorem}

Methods for finding piecewise-constant potentials 
from fixed-energy shifts were developed in
\cite{ars},
\cite{rs}, \cite{rsmi}. 
Numerical results show that, 
practically, the solution of inverse scattering problem  
with fixed-energy data can be very unstable towards
small perturbation of the data. 
There exist quite different
 potentials which produce practically the same phase
shifts (at one
energy level).
 The inverse scattering problem with fixed-energy data is 
 ill-posed, see \cite{ars} and \cite{rsmi}. 
After the publication of the example
in \cite{ars}, S. Gutman in \cite{gut5} has
constructed several examples of potentials having practically the same
phase shifts, and examined the dependency of the phase shifts on the
energy level $k$. Based on these observations, a best 
fit to data method
involving several energy levels has been proposed and implemented. 

However, the
potentials in the above examples change sign.
 Such potentials can be viewed by some
physicists as "less physical than one would like to". It has
been conjectured by these physicists that it is not
possible to construct two different positive potentials
which produce practically the same phase shifts at
a fixed energy and all angular momenta.
AGR has discussed this with Professor W. Scheid (Giessen,
Germany).
In this note we construct four positive piecewise-constant
spherically symmetric potentials having practically identical phase
shifts, thus proving that in fact such construction is possible contrary
to the belief of some physicists.

Phase shifts and an algorithm for their computation are discussed in
Section 2. A hybrid stochastic-deterministic method for finding
potentials with practically identical shifts is presented in Section 3.
Numerical experiments and the 
constructed potentials are described in Section 4.

\section{Phase Shifts for Piecewise-Constant Potentials}

Below a method from [1] is summarized.
Consider a finite set of points $0=r_0< r_1 <r_2<\dots<r_N=R$
and a piecewise-constant potential
\be\label{pot}
q(r)=q_i,\t{ on } [r_{i-1},r_i) \t{ for } 
i=1,\dots, N, \t{ and } q=0\t{ for }r\ge R.
\end{equation}

Denote $\k_i^2:=k^2-q_i$, where $i=1,\dots, N,$
and $k$ is some fixed positive number.
Consider the following problem for the radial Schr\"odinger equation:
\be
\f{d^2\p_l}{dr^2}+\Biggl(k^2-\f{l(l+1)}{r^2}\Biggl)\p_l=q\p_l,
\q \lim_{r\to 0}(2l+1)!!r^{-l-1}\p_l(r)=1,
\end{equation}

which we rewrite as:

\be\label{sc}
\f{d^2\p_l}{dr^2}+\Biggl(\k_i^2-\f{l(l+1)}{r^2}\Biggl)\p_l=0
\end{equation}

on the interval $r_{i-1}\le r < r_i$.
On $[r_{i-1},r_i)$ one has the following general solution of (\ref{sc})

\be
\p_l(r)=A_ij_l(\k_ir)+B_in_l(\k_ir).
\end{equation}

We assume below that $\k_i$ does not vanish for all $i$. If $\k_i=0$
for some $i$ then our approach is still valid with obvious changes.

>From the regularity of $\p_l$ at zero one gets $B_1=0$. Denote
$x_i=B_i/A_i$, then $x_1=0$. 
We are looking for the continuously differentiable solution $\p_l$.
Thus, the following interface conditions hold:
\be\ba{lcc}
A_ij_l(\k_ir_i)+B_in_l(\k_ir_i)=A_{i+1}j_l(\k_{i+1}r_{i})+
B_{i+1}n_l(\k_{i+1}r_{i}),\\ \\
\f{\k_i}{\k_{i+1}}[A_ij_l^\pr(\k_ir_i)+B_in_l^\pr(\k_ir_i)]=
A_{i+1}j_l^\pr(\k_{i+1}r_{i})+B_{i+1}n^\pr_l(\k_{i+1}r_{i}).
\end{array}
\end{equation}

The Wronskian $W(j_l(r),n_l(r))=1$, thus
\be\ba{lcc}
A_{i+1}=n^\pr_l(\k_{i+1}r_{i})[A_ij_l(\k_ir_i)+B_in_l(\k_ir_i)]
-\f{\k_i}{\k_{i+1}}n_l(\k_{i+1}r_{i})[A_ij_l^\pr(\k_ir_i)+B_in_l^\pr(\k_ir_i)],
\\
\\
B_{i+1}=\f{\k_i}{\k_{i+1}}j_l(\k_{i+1}r_{i})[A_ij_l^\pr(\k_ir_i)+
B_in_l^\pr(\k_ir_i)]
-j^\pr_l(\k_{i+1}r_{i})[A_ij_l(\k_ir_i)+B_in_l(\k_ir_i)].
\end{array}
\end{equation}
Therefore
\be
\begin{pmatrix}
A_{i+1}\\ B_{i+1}
\end{pmatrix}
=\f{1}{\k_{i+1}}
\begin{pmatrix}\a^i_{11} & \a^i_{12}\cr
\a^i_{21} & \a^i_{22}
\end{pmatrix}
\begin{pmatrix}A_{i}\cr B_{i}
\end{pmatrix},
\end{equation}
where the entries of the matrix $\alpha^i$ can be written explicitly:
\be\ba{lcc}
\a^i_{11}=\k_{i+1}j_l(\k_ir_i)n^\pr_l(\k_{i+1}r_{i})-
\k_{i}j_l^\pr(\k_ir_i)n_l(\k_{i+1}r_{i}),
\\
\\
\a^i_{12}=\k_{i+1}n_l(\k_ir_i)n^\pr_l(\k_{i+1}r_{i})-
\k_{i}n_l^\pr(\k_ir_i)n_l(\k_{i+1}r_{i}),
\\
\\
\a^i_{21}=\k_{i}j^\pr_l(\k_ir_i)j_l(\k_{i+1}r_{i})-
\k_{i+1}j_l(\k_ir_i)j_l^\pr(\k_{i+1}r_{i}),
\\
\\
\a^i_{22}=\k_{i}n_l^\pr(\k_{i}r_i)j_l(\k_{i+1}r_{i})-
\k_{i+1}n_l(\k_ir_i)j_l^\pr(\k_{i+1}r_{i}).
\ea
\end{equation}

Thus
\be\label{xk}
x_{i+1}=\f{\a^i_{21}+\a^i_{22}x_i}{\a^i_{11}+\a^i_{12}x_i},\q
x_i:=\f{B_i}{A_i}
\end{equation}

The phase shift $\d(k,l)$ is defined by

\be
\p_l(r)\sim{|F(k,l)|\over k^{l+1}}
\sin(kr-\frac{\pi l}{2}+
\delta(k,l))\quad r\to\infty\enspace ,
\end{equation}

where $F(k,l)$ is the Jost function.
For $r>R$

\be\label{as}
\p_l(r)=A_{N+1}j_l(kr)+B_{N+1}n_l(kr).
\end{equation}

>From (\ref{as}) and the asymptotics
$j_l(kr)\sim\sin(kr-l\pi/2),\q n_l(kr)\sim-\cos(kr-l\pi/2)$,
$r\to\infty$, one gets:

\be\label{dkl}
\tan\delta(k,l)=-\f{B_{N+1}}{A_{N+1}}=-x_{N+1}\,.
\end{equation}
Finally, the phase shifts of the potential $q(r)$ defined in (\ref{pot})
are found from
\be\label{dklf}
\delta(k,l)=-\arctan x_{N+1}.
\end{equation}

Let $q_0(r)$ be a spherically symmetric piecewise-constant potential.
Fix an energy level $k$ and a sufficiently large $N$. Let
$\{\tilde\delta(k,l)\}_{l=1}^N$ be the set of its phase shifts.
Let $q(r)$ be another such potential, and let
 $\{\delta(k,l)\}_{l=1}^N$ be the set of its phase shifts.

The best fit to data
function $\Phi(q,k)$ is defined by

\begin{equation}\label{phi}
\Phi(q,k)=\left(\frac{\sum^N_{l=1}\abs{\delta(k,l)-
\tilde\delta(k,l)}^2}
{\sum^N_{l=1}\abs{\tilde\delta(k,l)}^2}\right)^{1/2}\,,
\end{equation}

For sufficiently large $N$ such a function is practically the same as
the one which would use all the shifts in (\ref{phi}), since the phase
shifts are known to decay rapidly with $l$, see \cite{rai}.
Our goal is to find positive potentials $q(r)$, quite distinct from
$q_0(r)$, that make the objective function small. This is a complex nonlinear
minimization problem. An algorithm for its solution is given in the next
Section.

\section{  Global and Local Minimization Methods}

We seek the potentials $q(r)$ in the class of piecewise-constant, spherically 
symmetric real-valued functions. Let the admissible set be
\begin{equation}\label{adm}
A_{adm} \subset \{(r_1,r_2,\dots,r_M,q_1,q_2,\dots,q_M)\ : \ 0\leq r_i\leq R\,,\ 
q_{low}\leq q_m \leq q_{high}\}\,,
\end{equation}

where the bounds $q_{low}$ and $q_{high}$
for the potentials, as well as the bound $M$ on
the expected number of layers are assumed to be given.

A configuration $(r_1,r_2,\dots,r_M,q_1,q_2,\dots,q_M)$ corresponds to
the potential

\begin{equation}
q(r)=q_m\,,\quad \text{for}\quad r_{m-1}\leq r<r_m\,,\quad 1\leq m\leq M\,, 
\end{equation}
where $r_0=0$ and $q(r)=0$ for $r\geq r_M=R$.

Note, that the admissible configurations must also satisfy

\begin{equation}\label{admr}
r_1\leq r_2\leq r_3 \leq\dots\leq r_M\,.
\end{equation}

Given an initial configuration $Q_0\in A_{adm}\subset
\rc^{2M}$ a local minimization method finds a
local minimum near $Q_0$. On the other hand, global minimization methods
explore the entire admissible set to find a global minimum of the
objective function. While the local minimization is usually
deterministic, the majority of the global methods are probabilistic in
their nature. 
As usual for these type of inverse problems
the best fit to
data function $\Phi$ has many local minima, see \cite{gut5}. In this situation
it is exceedingly unlikely to get the minima points by chance alone.
Thus our special interest is for the minimization methods that combine
a global search with a local minimization. In \cite{gutmanramm} we 
developed such a method (the Hybrid Stochastic-Deterministic Method),
and applied it for the identification of small subsurface particles,
given a set of surface measurements. The HSD method could be
classified as a variation of a genetic algorithm with a local search
with reduction. In \cite{gut3}  
two global search algorithms in combination with
a special local search method were applied to the identification of 
piecewise-constant scatterers by 
acoustic type measurements. The global algorithms considered are
the Deep's Method, and Rinnooy Kan and Timmer's 
Multilevel Single-Linkage Method. 
The MSLM method has been applied to the identification of piecewise-constant
spherically symmetric potentials by their phase shifts in \cite{gut5}.

The potentials in this paper have been found using the so-called Reduced
Random Search Method.

In a pure {\bf Random Search} method a batch $H$ of $L$ trial points is generated in
$A_{adm}$ using a uniformly distributed random variable. Then a local
search is started from each of these $L$ points. A local minimum with
the smallest value of $\Phi$ is declared to be the global one.

A refinement of the Random Search is 
the {\bf Reduced Sample Random Search} method.
Here we use
only a certain fixed fraction $\gamma<1$ of the original batch of $L$
points to proceed with the local searches. This reduced sample $H_{red}$ of
$\gamma L$ points is chosen to contain the points with the smallest
$\gamma L$ values of $\Phi$ among the original batch. The local searches
are started from the points in this reduced sample.

While this method may not be as efficient as the full MSLM algorithm, it
has proved to be adequate for the problem at hand.

In our minimization algorithm the Reduced Random Search method is coupled with
a deterministic Local Minimization Method.

Numerical experience shows that the
objective function $\Phi$ is relatively well behaved in this problem.
While it contains many local minima and, at some points, $\Phi$ is not
differentiable, standard minimization methods work well here. 
A Newton type method for the minimization of $\Phi$ is described in
\cite{ars}. We have chosen to use a variation of Powell's minimization
method which does not require the computation of the derivatives of the
objective function. Such method needs a minimization routine for a one-
dimensional minimization of $\Phi$, which we do using a Bisection or a
Golden Rule method. See \cite{gut3} or \cite{gut5} for a complete
description. 

Now we can describe our Basic Local Minimization
Method in $\rc^{2M}$, which is a modification 
of Powell's minimization method \cite{bre}.

\subsection*{Basic Local Minimization Method}
\begin{enumerate}

\item  Initialize the set of directions $u_i$ to the basis vectors
\[
u_i=e_i\,,\quad i=1,2,\dots,2M\,.
\]

\item  Save your starting position as $Q_0$.
\item  For $i=1,\dots,2M$ move from $Q_0$ along the direction $u_i$ 
 and find the point of minimum $Q_i^t$.
\item  Re-index the directions $u_i$, 
so that (for the new indices) $\Phi(Q_1^t)\leq \Phi(Q_2^t)
\leq,\dots,\Phi(Q_{2M}^t)\leq\Phi(Q_0)$.

\item  For $i=1,\dots,2M$ move from 
$Q_{i-1}$ along the direction $u_i$ and find the point
of minimum $Q_i$.
\item  Set $v=Q_{2M}-Q_0$.
\item  Move from $Q_0$  along the direction $v$ and find the minimum. Call it
 $Q_0$. It replaces $Q_0$ from step 2.
\item Repeat the above steps untill a stopping criterion is satisfied.
\end{enumerate}

Note, that we use the temporary points of minima $Q_i^t$ only to rearrange the
initial directions $u_i$ in a different order.

Still another refinement of the local phase is necessary
to produce a successful minimization. The admissible set $A_{adm}$, see
(\ref{adm})-(\ref{admr}), belongs to a $2M$ 
dimensional minimization space $\rc^{2M}$.
The dimension $2M$ of this space
is chosen a priori to be larger than $2N$, where $N$
is the number of layers in the original potential. We have chosen
$M=6$ in our numerical experiments. 
However, since the sought potential may have fewer than 
$M$ layers, we found that conducting searches in lower-dimensional subspaces
of $\rc^{2M}$ is essential for the local minimization phase.
A variation of the following "reduction" procedure has also been found 
to be necessary in \cite{gutmanramm} for the search of small subsurface objects, 
and in \cite{gut3} for the 
identification of multilayered scatterers.

If two adjacent layers in a potential
have  values $v_{i-1}$ and $v_i$ and the objective function $\Phi$
is not changed much when both layers are 
assigned the same  value $v_i$ (or $v_{i-1}$),
 then these two layers can be replaced with just one occupying their place. 
The minimization problem becomes
constrained to a lower dimensional subspace of $\rc^{2M}$ and the local
minimization is done in this subspace.

\subsection*{Reduction Procedure}

Let $\epsilon_r$ be a positive number.

\begin{enumerate}

\item  Save your starting configuration
$Q_0=(r_1,r_2,\dots,r_M,v_1,v_2,\dots,v_M)\in A_{adm}$
 and the value $\Phi(Q_0)$. Let the $M+1$-st 
layer be $D_{M+1}=\{r_M\leq |x| \leq
R\}$ and $v_{M+1}=0$.

\item  For $i=2,\dots,M+1$ replace $v_{i-1}$ 
in the layer $D_{i-1}$ by $v_i$. Compute
$\Phi$ at the new configuration  $Q_i^d$, and the difference
$c_i^d=|\Phi(Q_0)-\Phi(Q_i^d)|$.

\item  For $i=1,\dots,M$ replace $v_{i+1}$ 
in the layer $D_{i+1}$ by $v_i$. Compute
$\Phi$ at the new configuration  $Q_i^u$, and the difference
$c_i^u=|\Phi(Q_0)-\Phi(Q_i^u)|$.

\item  Find the smallest among the numbers $c_i^d$ and $c_i^u$.
If this number is less than $\epsilon_r\Phi(Q_0)$, then adjust the
value of the potential to $v_i$ in the "down" or "up" layer accordingly.
Replace the two adjacent layers with one occupying their place, and
renumber the layers.

\item Repeat the above steps untill no further reduction in the number of
layers is occurring.

\end{enumerate}

Note, that an application of the Reduction Procedure may or may not
result in the actual reduction of layers.

Finally, the entire Local Minimization Method {\bf (LMM)} consists of the
following:

\subsection*{Local Minimization Method (LMM)}
\begin{enumerate}

\item  Let your starting configuration be
$Q_0=(r_1,r_2,\dots,r_M,v_1,v_2,\dots,v_M)\in A_{adm}$.

\item  Apply the Reduction Procedure to $Q_0$, 
and obtain a reduced configuration
$Q_0^r$ containing $M^r$ layers.

\item  Apply the Basic Minimization Method in $A_{adm}\bigcap \rc^{2M^r}$ 
with the starting point $Q_0^r$, and obtain a configuration $Q_1$.

\item  Apply the Reduction Procedure to 
$Q_1$, and obtain a final reduced configuration
$Q_1^r$.

\end{enumerate}

\section{Numerical Results}
Let $q_0$ be the following potential
\[
q_0(r)=\begin{cases} 
7.2 & 0\leq r < 0.5\\
4.5 & 0.5 \leq r < 1.0\\
7.2 & 1.0 \leq r < 1.5\\
4.5 & 1.5 \leq r < 2.0\\
0.0 &  r \geq 2.0
\end{cases}
\]


\begin{table}
\caption{Phase shifts of $q_0(r)$ for $k=3$.}

\begin{tabular}{r r| r r| r r}

\hline

$l$ & $\tilde\delta(k,l)$ & $l$ & $\tilde\delta(k,l)$ & $l$ &
$\tilde\delta(k,l)$ \\
\hline

 0 & -0.220024E+00 & 7 & -0.183339E-02 & 14 & -0.204010E-10 \\
 1 & -0.188623E+00 & 8 & -0.250850E-03 & 15 &  -0.766553E-12 \\ 
 2 & -0.210693E+00 & 9 & -0.267137E-04 & 16 &  -0.253238E-13 \\
 3 & -0.185306E+00 & 10 & -0.228367E-05 & 17 &  -0.741554E-15 \\
 4 & -0.104318E+00 & 11 & -0.160476E-06 & 18 & -0.193858E-16 \\
 5 & -0.390310E-01 & 12 & -0.944572E-08 & 19 & -0.455299E-18 \\
 6 & -0.100159E-01 & 13 & -0.472923E-09 & 20 & -0.966113E-20 \\
 
\hline
\end{tabular}

\end{table} 
Let $k=3$. The phase shifts $\tilde\delta(k,l)$ are computed as in
Section 2. They are shown in Table 1.
Given another piecewise-constant spherically symmetric potential $q(r)$,
its phase shifts $\delta(k,l)$ 
are computed in the same way. The objective function $\Phi(q)$
is formed according to (\ref{phi}) with $N=20$. 
This objective function is minimized over 
$A_{adm}$ (\ref{adm})  with $M=6$ and $R=3.0$ as described in Section 3. 
A priori bounds for the potential were chosen to be
$q_{low}=0.0$ and $q_{high}=9.0$. The Reduced Random Search with the
batch size $L=10000$ and the reduction factor $\gamma=0.01$ was used for
the global minimization. The value $\epsilon_r=0.1$ was used in
the Reduction Procedure (see Section 3) during the local minimization phase.
The initial configurations were generated using a random 
number generator with seeds determined by the system time.
The run time was about 30 minutes on a 333 MHz PC.

Three potentials with $\Phi(q)<10^{-4}$ obtained by this procedure are,
for example,
\[
q_1(r)=\begin{cases} 
8.9991 & 0\leq r < 0.4316\\
3.9672 & 0.4316 \leq r < 0.8758\\
6.7356 & 0.8758 \leq r < 1.5718\\
4.3029 & 1.5718 \leq r < 2.0065\\
0.0 &  r \geq 2.0065
\end{cases}
\]
with $\Phi(q_1)=9.3586605\cdot 10^{-5}$, 
\[
q_2(r)=\begin{cases} 
6.4197 & 0\leq r < 0.6809\\
3.1509 & 0.6809 \leq r < 0.9162\\
6.7464 & 0.9162 \leq r < 1.2856\\
8.7210 & 1.2856 \leq r < 1.4314\\
4.5936 & 1.4314 \leq r < 1.9969\\
0.0 &  r \geq 1.9969
\end{cases}
\]
with $\Phi(q_2)=6.1848208\cdot 10^{-5}$, and
\[
q_3(r)=\begin{cases} 
5.9463 & 0\leq r < 0.8666\\
0.1008 & 0.8666 \leq r < 0.9862\\
7.9164 & 0.9862 \leq r < 1.4345\\
4.6116 & 1.4345 \leq r < 1.9964\\
0.0 &  r \geq 1.9964
\end{cases}
\]
with $\Phi(q_3)=3.3089927\cdot 10^{-5}$.

Figures 1, 2 and 3 show these potentials (solid lines) as
well as the original potential $q_0(r)$. 
Potentials  $q_1\,,\:q_2$ and $q_3$ are quite
different from $q_0$, but their phase shifts are practically identical
to the phase shifts of $q_0$. Since real measurements always contain
some noise, such a potential is impossible to identify from its phase
shifts (at this energy level)
by any inverse method.

\begin{figure}[fig1]
\vspace{5pc}
\includegraphics*{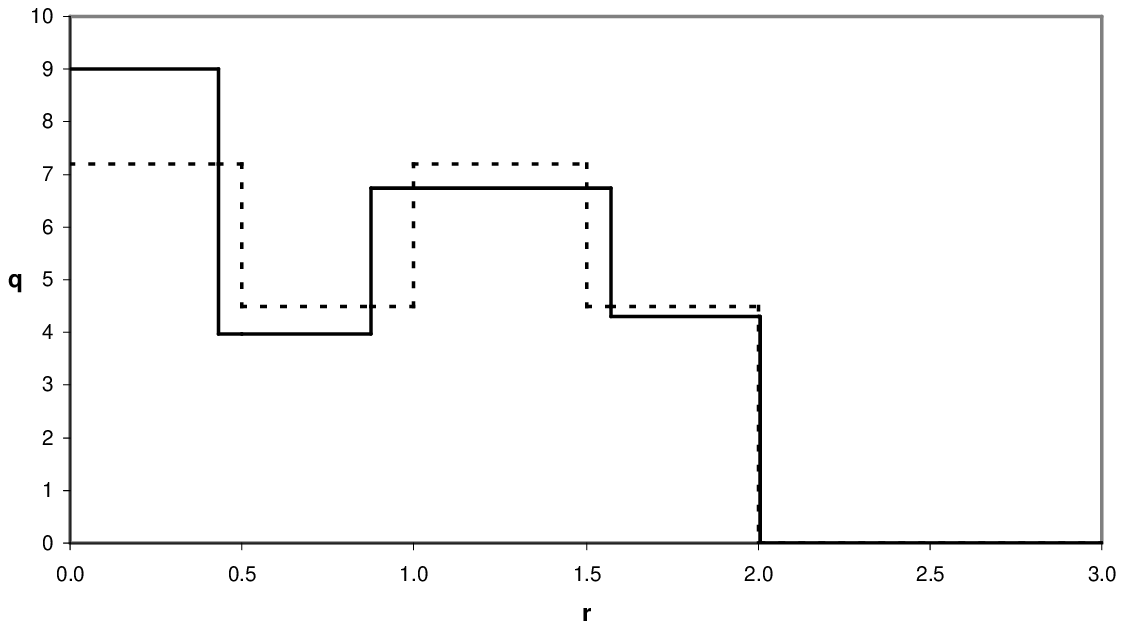}
\caption{Potential $q_1(r)$ (solid line), and the original
potential $q_0(r)$ (dotted line);
$\Phi(q_1)=9.3586605\cdot 10^{-5}$.}
\end{figure}

\begin{figure}[fig2]
\vspace{5pc}
\includegraphics*{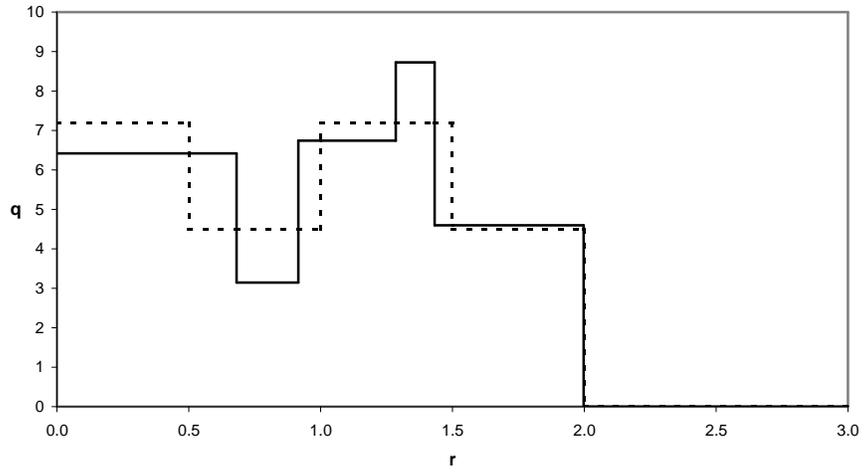}
\caption{Potential $q_2(r)$ (solid line), and the original
potential $q_0(r)$ (dotted line);
$\Phi(q_2)=6.1848208\cdot 10^{-5}$.}
\end{figure}

\begin{figure}[fig3]
\vspace{5pc}
\includegraphics*{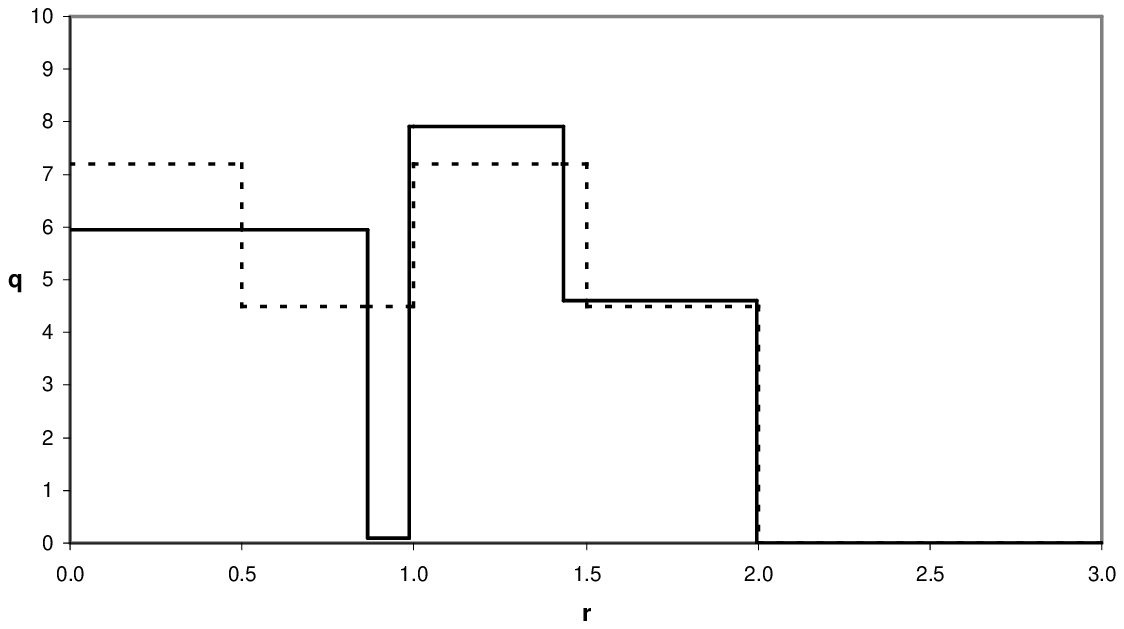}
\caption{Potential $q_3(r)$ (solid line), and the original
potential $q_0(r)$ (dotted line);
$\Phi(q_3)=3.3089927\cdot 10^{-5}$.}
\end{figure}

\section{Conclusions}
Recovery of a spherically symmetric potential from its
fixed-energy phase
shifts is an important physical problem. Recent theoretical results (see
Theorem 1.1) assure that such a potential is uniquely defined by a sufficiently
large subset of its phase shifts at any one fixed energy level. However,
two different potentials can produce almost identical fixed-energy phase
shifts. Such examples were obtained in \cite{ars} and \cite{gut5}.
These examples give oscillating sign-changing potentials.
It was suggested  by some physicists
that such potentials are "less physical". 
In this note we present
examples of positive spherically symmetric 
piecewise-constant finite-range potentials having practically the same
phase shifts (at a fixed energy level). 
The existence of such potentials shows that 
the inverse scattering problem with fixed-energy data
can be very ill-posed even if the enrgy level is not too small.
 The solution of the inverse scattering problem
becomes much more stable if
the phase shifts are known at several energy levels, see
\cite{gut5} for details. It is of interest to study the 
stability
of the solution to inverse scattering problem with
fixed-energy phase shifts 
if one assumes a priori that the potential is not only positive but
also smooth.

\medskip
{\bf Acknowledgement.} AGR thanks Professor W. Scheid 
for stimulating discussions.

\end{document}